\documentclass[apj,tighten,numberedappendix,appendixfloats]{emulateapj}
\usepackage{natbib}
\usepackage{color}

\newcommand{\nhcen}{n_{\rm H,cen}}
\newcommand{\cc}{{\rm cm^{-3}}}

\newcommand{\msun}{M_\odot}
\newcommand{\msunyr}{M_\odot~{\rm yr}^{-1}}

\shorttitle{Primordial Power Spectrum and the First Stars}
\shortauthors{Hirano et al.}

\begin{document}

\title{Early structure formation from primordial density fluctuations\\
with a blue-tilted power spectrum}

\author{
Shingo Hirano\altaffilmark{1},
Nick Zhu\altaffilmark{1,3},
Naoki Yoshida\altaffilmark{1,2},
David Spergel\altaffilmark{3},
and
Harold W. Yorke\altaffilmark{4}
}

\altaffiltext{1}{Department of Physics, 
University of Tokyo, Bunkyo, Tokyo 113-0033, Japan}
\altaffiltext{2}{Kavli Institute for the Physics and Mathematics 
of the Universe (WPI), Institutes for Advanced Study, 
University of Tokyo, Kashiwa, Chiba 277-8583, Japan}
\altaffiltext{3}{Department of Astrophysical Sciences,
Princeton University, Peyton Hall, Princeton, NJ 08544, USA}
\altaffiltext{4}{Jet Propulsion Laboratory, 
California Institute of Technology, Pasadena, CA 91109, USA}

\setcounter{footnote}{-1}

\begin{abstract}
While observations of large-scale structure and the cosmic microwave background (CMB) provide strong constraints on the amplitude of the primordial power spectrum (PPS) on scales larger than 10~Mpc, the amplitude of the power spectrum on sub-galactic length scales is much more poorly constrained. 
We study early structure formation in a cosmological model with a blue-tilted PPS.
We assume that the standard scale-invariant PPS is modified at small length scales as $P(k) \sim k^{m_{\rm s}}$ with $m_{\rm s} > 1$.
We run a series of cosmological hydrodynamic simulations to examine the dependence of the formation epoch and the characteristic mass of primordial stars on the tilt of the PPS. 
In models with $m_{\rm s} > 1$, star-forming gas clouds are formed at $z > 100$, when formation of hydrogen molecules is inefficient because the intense CMB radiation destroys chemical intermediates.
Without efficient coolant, the gas clouds gravitationally contract while keeping a high temperature.
The protostars formed in such ``hot'' clouds grow very rapidly by accretion to become extremely massive stars that may leave massive black holes with a few hundred solar-masses at $z > 100$.
The shape of the PPS critically affects the properties and the formation epoch of the first generation of stars.
Future experiments of the CMB polarization and the spectrum distortion may provide important information on the nature of the first stars and their formation epoch, and hence on the shape of the small-scale power spectrum.
\end{abstract}


\section{Introduction}

The formation of the first stars marks the end of the cosmic Dark Ages, when their energetic photons initiates reionization of the inter-galactic medium. 
The exact epoch when cosmic reionization began is still uncertain, but evidence is mounting from an array of observations that the first luminous objects such as galaxies and quasars were formed very early on \citep{watson15,wu15,planck15XIII}. 
\cite{wu15} recently reported the discovery of a supermassive black hole in an extremely bright quasar. 
While their exact estimate of a black hole mass of twelve billion solar masses rests on the controversial use of broad emission lines \citep[see e.g.,][]{cackett15}, this ultra-luminous quasar must host a supermassive black hole.  
There are now over 40 quasars known with $z > 6$ including ULAS J1120+0641 with a $2 \times 10^9~\msun$ black hole at $z = 7.085$ \citep{mortlock11,derosa14} and a recently discovered $z = 6.889$ quasar with a black hole of $2.1 \times 10^9~\msun$ \citep{derosa14}. 
Assembling these giant quasar hosts by this early epoch poses a serious challenge to the theory of structure formation and the growth of black holes.

There has been significant progress in the past decade in theoretical studies of the formation of the first generation of stars \citep[see][for recent reviews]{bromm13,glover13,greif15}.
Owing largely to the fact that the initial conditions are cosmologically well determined, one can perform {\it ab initio} simulations of early structure formation \citep[e.g.,][]{bromm02,abel02,yoshida03}. 
An important element of such simulations is the primordial power spectrum (PPS) from which a realistic initial density field can be generated in a fully cosmological context.

The PPS is observationally determined to have a power-law with a spectral index $n_{\rm s} \simeq 0.96$ \citep{planck15XIII}.
Unfortunately, such tight constraints are derived only at large length scales, up to the wavenumber of $k \simeq 0.2~{\rm Mpc}^{-1}$ \citep{hlozek12}.  
Most of theoretical studies thus assume a scale-invariant PPS and adopt significant extrapolation of the observationally determined large-scale power spectrum. 
The observational evidence on sub-galactic scales is confusing: milli-lensing measurements suggest substructure consistent with the extrapolation of the $\Lambda$-Cold Dark Matter ($\Lambda$-CDM) power spectrum to smaller scales \citep{hezaveh13,bussmann13,xu15} yet studies of local dwarf galaxies suggest significant problems with $\Lambda$-CDM on small scales \citep{strigari07}.
It is not only interesting to explore how early structure formation is affected by enhanced density fluctuations at small scales, but also timely because the existence of supermassive black holes at very high redshifts suggests rapid growth of small scale structure in the early universe.

A variety of possibilities are proposed from the physics of the early universe that posit scale-dependent PPS.
Popular single-field inflation models predict a nearly scale-free and slightly red power spectrum for adiabatic scalar perturbations. 
There are also variant models that predict scale-dependent power spectrum with kinks, bumps, and other types of features \citep[e.g.,][]{starobinskii92,adams97,kawaguchi08,biswas10,barnaby10}.
Of particular interest to us in the present paper are the models that generate density perturbations with a blue-tilted power spectrum or with enhancement at small length scales \citep{covy99,martin01,gong11}.

\begin{figure}
\begin{center}
\resizebox{7.5cm}{!}{\includegraphics[clip,scale=1]{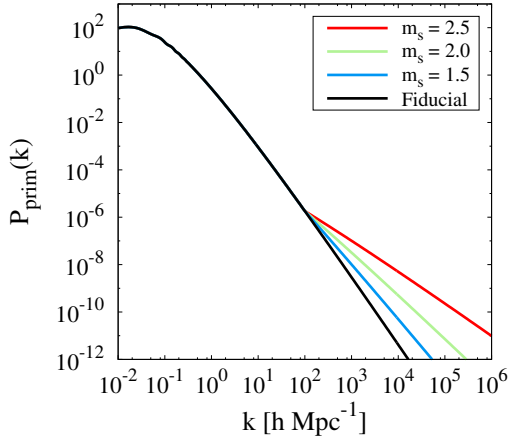}}
\caption{
We plot the matter power spectra for the considered cases in the present paper with $k_{\rm p} = 100~h~{\rm Mpc}^{-1}$. 
The black line is for the scale-independent PPS (Eq.~\ref{eq:N-PPS}) whereas the other three lines show the scale-dependent models given by Eq.~\ref{eq:E-PPS} with $m_{\rm s}$ = 1.5 (blue), 2.0 (green), and 2.5 (red), respectively.
}
\label{fig:PPS}
\end{center}
\end{figure}

It is naively expected that structure forms early in such models, yielding abundant small mass dark halos. 
Because primordial star formation involves a number of physical processes such as chemistry and radiative transfer
as well as gravitational assembly of dark halos, the initial density perturbations or the PPS can affect the formation
of primordial stars in a complicated manner. 
Clearly, it is important to study the properties of the first structure using cosmological simulations with the relevant physics included.

In the present paper, we perform a series of cosmological simulations for different PPS models to study early structure formation in detail.
In particular, we examine how the slope of the PPS at the small length scales affects the formation epoch and the mass of the first generation of stars.
We begin by describing the calculation methods in Section~\ref{sec:method}.
Section~\ref{sec:res} shows the simulation results of primordial star formation for different PPS models. 
We discuss the dependence of the primordial star formation on the PPS model in Section~\ref{sec:dis}. 
Throughout this paper, we adopt the standard $\Lambda$-CDM cosmology with the total matter density $\Omega_{\rm m} = 0.3086$, the baryon density $\Omega_{\rm b} = 0.04825$, the dark energy density $\Omega_{\Lambda} = 0.6914$ in units of the critical density, the Hubble constant $h = 0.6777$, and the primordial index $n_{\rm s} = 0.9611$ \citep{PLANCK13XVI}.

\section{Numerical simulations}
\label{sec:method}

\begin{figure}
\begin{center}
\resizebox{7.5cm}{!}{\includegraphics[clip,scale=1]{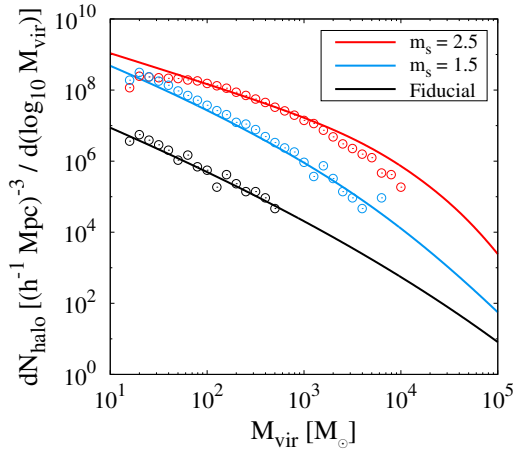}}
\caption{
We compare the halo mass functions at $z = 100$ for three PPS models.
We set the normalization parameter $\sigma_8 = 2.5$ here. 
The lines show the Press-Schechter mass functions, while the symbols show the results of our cosmological simulations with a volume of $60~h^{-1}$ comoving kilo-parsec on a side.
We run the Friend-Of Friend (FOF) halo finder with linking length $b = 0.2$ to obtain the halo mass functions.
}
\label{fig:HMF}
\end{center}
\end{figure}

We perform a series of cosmological simulations that start from the initial conditions generated for different PPS models $P_{\rm prim}(k)$. 
In the next sections, we describe the numerical methods with emphasis on the improvement over our earlier numerical simulations \citep{hirano14,hirano15}.

\subsection{Initial Conditions}

\begin{deluxetable*}{cccccrcrcccc}
\tablecaption{Properties of First Star Formation Depending on the PPS Models}

\tablehead{
\colhead{ID} & 
\colhead{$\sigma_8$} & 
\colhead{$k_{\rm p}$} & 
\colhead{$m_{\rm s}$} & 
\colhead{\ \ \ } & 
\colhead{$z_{\rm form}$} & 
\colhead{$M_{\rm vir}$} & 
\colhead{$M_{\rm Jeans}$} & 
\colhead{$\dot{M}_{\rm Jeans}$} & 
\colhead{\ \ \ } & 
\colhead{$M_{\rm *,cal}$} &
\colhead{$M_{\rm *,est}$} \\
\colhead{} & 
\colhead{} & 
\colhead{$(h~{\rm Mpc}^{-1})$} & 
\colhead{} & 
\colhead{} & 
\colhead{} & 
\colhead{$(10^5~\msun)$} & 
\colhead{$(\msun)$} & 
\colhead{$(10^{-3}~\msunyr)$} & 
\colhead{} & 
\colhead{$(\msun)$} & 
\colhead{$(\msun)$}
}

\startdata
100 & 1.5 & --  & --  & &  24.0 & 2.17 &  45 & 0.63 &  & \hspace{0.5mm} 45 & \hspace{0.5mm} 88 \\
111 &     & 100 & 1.5 & &  27.1 & 1.82 & 162 & 2.97 &  &     & 261 \\
112 &     &     & 2.0 & &  36.1 & 0.88 & 162 & 1.16 &  &     & 135 \\
113 &     &     & 2.5 & &  87.9 & 0.31 &  69 & 0.77 &  & \hspace{0.5mm} 96 & 101 \\
131 &     & 300 & 1.5 & &  24.0 & 2.11 &  23 & 0.24 &  &     & \hspace{0.5mm} 45 \\
132 &     &     & 2.0 & &  25.2 & 2.19 & 124 & 0.74 &  &     & \hspace{0.5mm} 98 \\
133 &     &     & 2.5 & &  33.9 & 1.97 & 171 & 3.31 &  &     & 281 \\
151 &     & 500 & 1.5 & &  23.5 & 2.10 &  33 & 0.35 &  &     & \hspace{0.5mm} 58 \\
152 &     &     & 2.0 & &  24.0 & 2.19 &  61 & 1.02 &  &     & 123 \\
153 &     &     & 2.5 & &  27.1 & 2.83 &  90 & 1.61 &  &     & 170 \\
\\                                                             
200 & 2.0 & --  & --  & &  34.9 & 1.30 & 401 & 1.54 &  & 114 & 165 \\
211 &     & 100 & 1.5 & &  40.1 & 0.83 & 167 & 1.29 &  &     & 145 \\
212 &     &     & 2.0 & &  53.2 & 0.22 &  57 & 0.25 &  &     & \hspace{0.5mm} 46 \\
213 &     &     & 2.5 & & 138.2 & 0.24 & 142 & 1.71 &  & 145 & 177 \\
\\                                                             
300 & 2.5 & --  & --  & &  46.2 & 0.83 & 243 & 2.34 &  & 118 & 220 \\
311 &     & 100 & 1.5 & &  52.6 & 0.65 & 164 & 1.39 &  &     & 153 \\
312 &     &     & 2.0 & &  72.1 & 0.22 &  60 & 0.43 &  &     & \hspace{0.5mm} 67 \\
313 &     &     & 2.5 & & 186.3 & 0.37 & 411 & 2.72 &  & 362 & 245 
\enddata

\tablecomments{
Column 1: Cloud index,
Column 2 -- 4: PPS model parameters,
Column 5 -- 8: Properties of star-forming clouds, 
Column 9: Stellar mass calculated by RHD simulations, and
Column 10: Stellar mass estimated by a fitting formula (\ref{eq:MIII_dMdt-Jeans}). 
The cloud ID indicates the combination of the three parameters $\{\sigma_8,~k_{\rm p},~m_{\rm s}\}$.
}

\label{tab:list}
\end{deluxetable*}

The standard scale-invariant PPS is given by
\begin{eqnarray}
P_{\rm prim}(k) \propto & k^{n_{\rm s}}~,
\label{eq:N-PPS}
\end{eqnarray} 
whereas the model power spectrum with enhancement at small scales is given by 
\begin{eqnarray}
P_{\rm prim}(k) & \propto & k^{n_{\rm s}} \hspace{15mm} (k < k_{\rm p})~, \nonumber \\
& \propto & k_{\rm p}^{n_{\rm s} - m_{\rm s}} \cdot k^{m_{\rm s}}~(k > k_{\rm p})~.  
\label{eq:E-PPS}
\end{eqnarray}
We adopt two parameters that characterize the spectrum; $k_{\rm p}$ is the pivot wavenumber above which the spectral index is set to be $m_{\rm s}$.
We consider $k_{\rm p} = $ 100, 300, and 500 $h~{\rm Mpc}^{-1}$ and $m_{\rm s} =$ 1.5, 2, and 2.5.

Figure~\ref{fig:PPS} shows the matter power spectra for models with $k_{\rm p} = 100~h~{\rm Mpc}^{-1}$.
The fiducial pure power-law PPS case is also shown there for comparison. 
We use the transfer function of \cite{eisenstein99} for this plot and also for generating the initial conditions using the publicly available code MUSIC \citep{hahn11}. 
The density fluctuation amplitude is normalized by $\sigma_8$.
Because our simulation volume is very small, we set $\sigma_8$ larger than the current standard value of about 0.8 \citep{planck15XIII}, so that we can simulate the formation of high-$\sigma$ peak halos. 
In practice, we set $\sigma_8 =$ 1.5, 2, and 2.5. 
In total, we generate 18 cosmological initial conditions with the parameter sets given in Table~\ref{tab:list}. 

The effect of the enhanced small-scale power is clearly seen in the mass function (Fig.~\ref{fig:HMF}).
The halo mass functions differ by orders of magnitude at $z = 100$.
Figure~\ref{fig:HMF} also shows that the Press-Schechter mass function agrees fairly well with the simulation results \citep[see also][]{reed05}.

\subsection{Cosmological Simulations}

We perform the cosmological simulations using the parallel N-body / Smoothed Particle Hydrodynamics (SPH) code Gadget-2 \citep{springel05} suitably modified for the primordial star formation simulations \citep{yoshida06,yoshida07,hirano14,hirano15}. 
In order to achieve sufficient numerical resolution, we use hierarchical zoom initial conditions. 
The parent simulation volume is $L_{\rm box} = 100~h^{-1}~{\rm kpc}$ on a side. 
The mass of the refined dark matter particles in the zoomed region is 0.80~$\msun$, and that of the baryonic component is 0.15~$\msun$.
We follow structure formation from $z_{\rm ini} = 499$ until the central hydrogen number density reaches $\nhcen \simeq 10^7~\cc$ or $10^{13}~\cc$ in a gas cloud that first collapses (see Section \ref{sec:method_mass}). 
The initial ionization fraction is computed by using RECFAST \citep{seager99,seager00,wong08} as $x_{\rm e} = 6.88 \times 10^{-4}$ at the initial redshift. 
During the cloud collapse, we keep the refinement criterion that the local Jeans length is always resolved by 15 times the local smoothing length by progressively increasing the spatial resolution by the particle-splitting technique \citep{kitsionas02}.

Let us briefly describe the overall trend found from the simulation result.
In the standard model, the first stars are formed at $z \sim 20 - 50$ \citep[e.g.,][]{tegmark97,reed05,yoshida03}. 
In one of our simulations with $\sigma_8 = 1.5$, the first star is formed at $z_{\rm form} = 24$ within the simulation volume of 100 $h^{-1}$~kpc on a side. 
We label this sample as ID 100. 
Here, the first digit ``1'' indicates the value of $\sigma_8$ (see Table~\ref{tab:list}).
We also find that, by changing $\sigma_8$ to 2.0 and 2.5 (ID = 200 and 300), the first star in the simulation volume is formed earlier at $z_{\rm form} = 35$ and 46, respectively.

\subsubsection{H$_2$ Formation in the Early Universe}
\label{sec:method_cos_h2}

We are interested in the thermal evolution of a primordial gas cloud formed at $z > 100$.
Because CMB photons at such high redshifts are capable of dissociating molecules and ions, several chemical reactions need to be included in order to follow the thermal and chemical evolution accurately.

At $z < 100$, the main H$_2$ formation path is the so-called H$^-$ channel; 
\begin{eqnarray}
{\rm H} + {\rm e}^- & \to & {\rm H}^- + h\nu~, \\
{\rm H}^- + {\rm H} & \to & {\rm H_2} + {\rm e}^-~.
\label{eq:HM_channel}
\end{eqnarray}
Because H$^-$ ions are destroyed by CMB photons at $z > 100$ via the 
photo-detachment reaction
\begin{eqnarray}
{\rm H}^- + h\nu \to {\rm H} + {\rm e}^-~,
\label{eq:HM_pd}
\end{eqnarray}
the H$^-$ channel is not effective at such early epochs.
An alternative path is the H$_2^+$ channel:
\begin{eqnarray}
{\rm H}^+ + {\rm H} & \to & {\rm H}_2^+ + h\nu~, \\
{\rm H}_2^+ + {\rm H} & \to & {\rm H}_2 + {\rm H}^+~.
\label{eq:H2+_channel}
\end{eqnarray}
An opposing dissociative process by energetic photons operates as 
\begin{eqnarray}
{\rm H}_2^+ + h\nu \to {\rm H}^+ + {\rm H}~.
\label{eq:H2+_pd}
\end{eqnarray}
At $z > 120$, most of the hydrogen molecules are formed by the H$_2^+$ channel. 
The mean H$_2$ fraction in the Universe reaches $\sim 10^{-6}$ at $z \sim 60$.
We update the chemical network by introducing the above reactions 
\citep[see a review by][]{galli13}. 
The reaction rates are based on \cite{coppola13}.
We describe more details of the chemistry implementation in Appendix~\ref{app:H2+}.

\begin{figure*}
\begin{center}
\resizebox{13.75cm}{!}{\includegraphics[clip,scale=1]{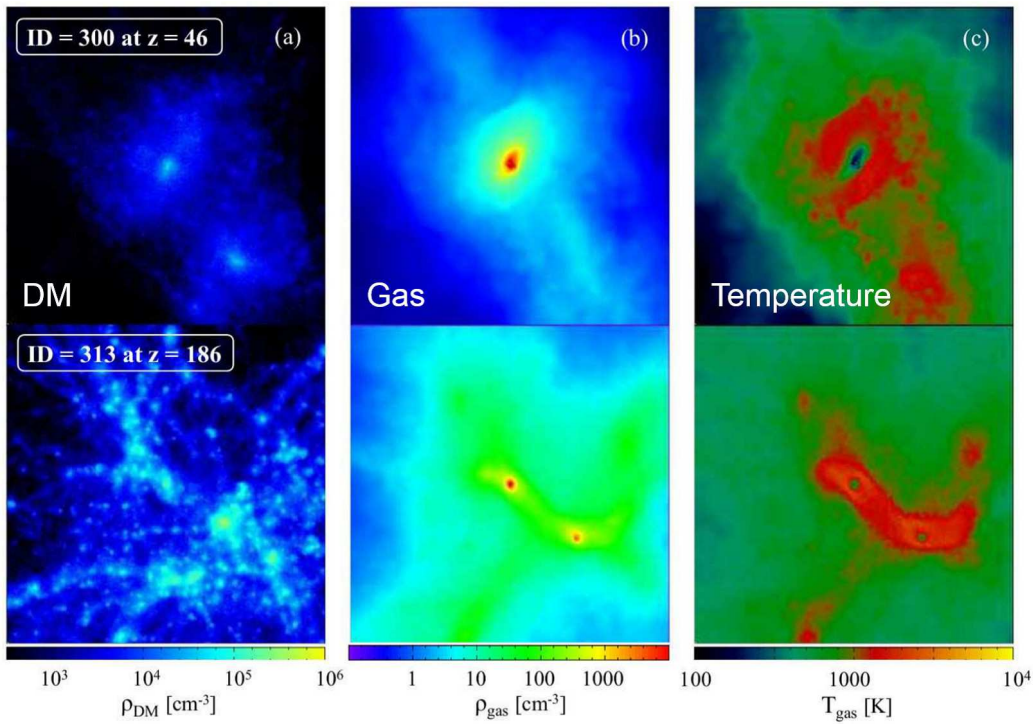}}
\caption{
Projected densities for dark matter (Panel a) and baryonic components (b). 
We also plot the projected gas temperature (c) at 
the formation sites of the primordial stars ID 300 (top; $z_{\rm form} = 46$) and ID 313 (bottom; $z_{\rm form} = 186$). The plotted region has a proper 100~pc on a side. We scale the density in units of particle number density 
per cubic-centimeter.
}
\label{fig:2dmap}
\end{center}
\end{figure*}

\subsection{Final Stellar Masses}
\label{sec:method_mass}

It is important to study the characteristic mass of the first stars in our variant cosmological models with blue-tilted PPS.
We calculate the mass of a primordial star by using two different methods. 
One is direct calculation of gas accretion onto the central protostar using the two-dimensional axisymmetric radiation hydrodynamic (RHD) simulation coupled with the stellar evolution \citep{hosokawa11,hosokawa12b,hirano14,hirano15}. 
We perform the calculations for 6 cases. 
We label them as ID = 100, 113, 200, 213, 300, and 313. 
Here, the second digit of the object ID indicates the pivot wavenumber, and the last digit indicates the value of ms (see Table~\ref{tab:list}).
For each gas cloud, we generate the initial conditions for the RHD simulation by suitably mapping the final snapshot of our cosmological simulation at the time when the density of the collapsing cloud core reaches $\nhcen = 10^{13}~\cc$. 
The RHD simulations are run until the mass accretion rate onto the central star falls below $10^{-4}~\msunyr$. 
At this point the stellar growth is almost completely halted and the resultant mass can be regarded as the final stellar mass. 
We denote the stellar masses calculated directly by the RHD simulation as $M_{\rm *,cal}$.

We also estimate the final stellar masses by using a fitting function that is derived from a number of RHD simulations of accreting protostars in our earlier studies.
The fitting function essentially correlates the stellar mass $M_{\rm *,est}$ to the gas infall rate $\dot{M}_{\rm Jeans}$ evaluated for a gravitationally Jeans unstable gas cloud. 
It is given by \cite{hirano15} as 
\begin{eqnarray}
M_{\rm *,est} = 250~\left( \frac{\dot{M}_{\rm Jeans}}{2.8 \times 10^{-3}~\msunyr} \right)^{0.7}~\msun~.
\label{eq:MIII_dMdt-Jeans}
\end{eqnarray}
By adopting this formula, we can skip the numerically costly RHD simulation in the accretion phase.
Note that the difference between the calculated and estimated values is about fifty percent, and mostly within a factor of two \citep[see fig.~14b in][]{hirano14}. 
Although this might still appear insufficient to determine the stellar mass for all the cases in Table~\ref{tab:list}, the mass variations caused by model parameters ($k_{\rm p}, m_{\rm s}$, etc) are typically larger than the uncertainty in the estimated mass (see Table~\ref{tab:list}).
Because we aim at investigating how cosmological parameters such as the slope of the power spectrum affect the formation of early halos and gas clouds, it would be clearer to interpret the resulting stellar masses in terms of gas clouds' properties such as the Jeans mass and accretion rate.
Therefore, we use the estimated stellar masses (Eq.~[\ref{eq:MIII_dMdt-Jeans}])  for all the 18 cases to discuss the qualitative dependence on the adopted PPS models.

Our stellar mass estimates are based on the results of radiation-hydrodynamics simulations which follow only the evolution of the central star.
Recent three-dimensional simulations show that the circumstellar disk can gravitationally fragment to produce multiple clumps \citep[sink particles; e.g.,][]{clark11,greif11,greif12,stacy13b}. 
Such fragments, if some of them survive over a long period of disk accretion, would form multiple stars with correspondingly small masses.
High-resolution simulations show also that fragments efficiently merge onto the central star in roughly an orbital timescale \citep[e.g.,][]{greif12,vorobyov13}.
In the latter case, the final stellar mass would be similar with the calculated value by our radiation-hydrodynamics simulation.
Conservatively, we interpret the derived $M_*$ as the total mass of stars formed, rather than the mass of a single star.

\begin{figure*}
\begin{center}
\resizebox{15cm}{!}{\begin{tabular}{cc}
\includegraphics[clip,scale=1]{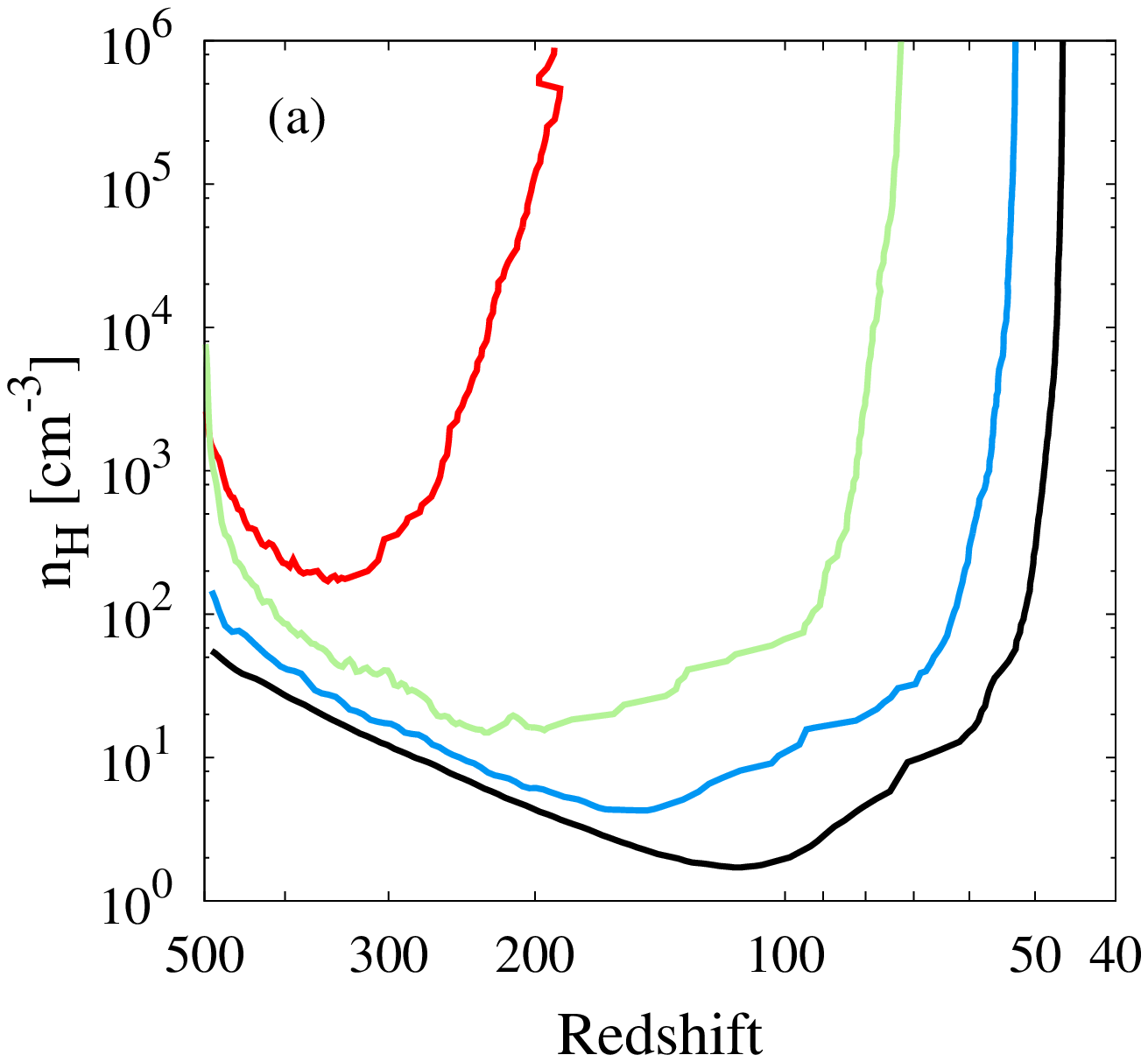}&
\includegraphics[clip,scale=1]{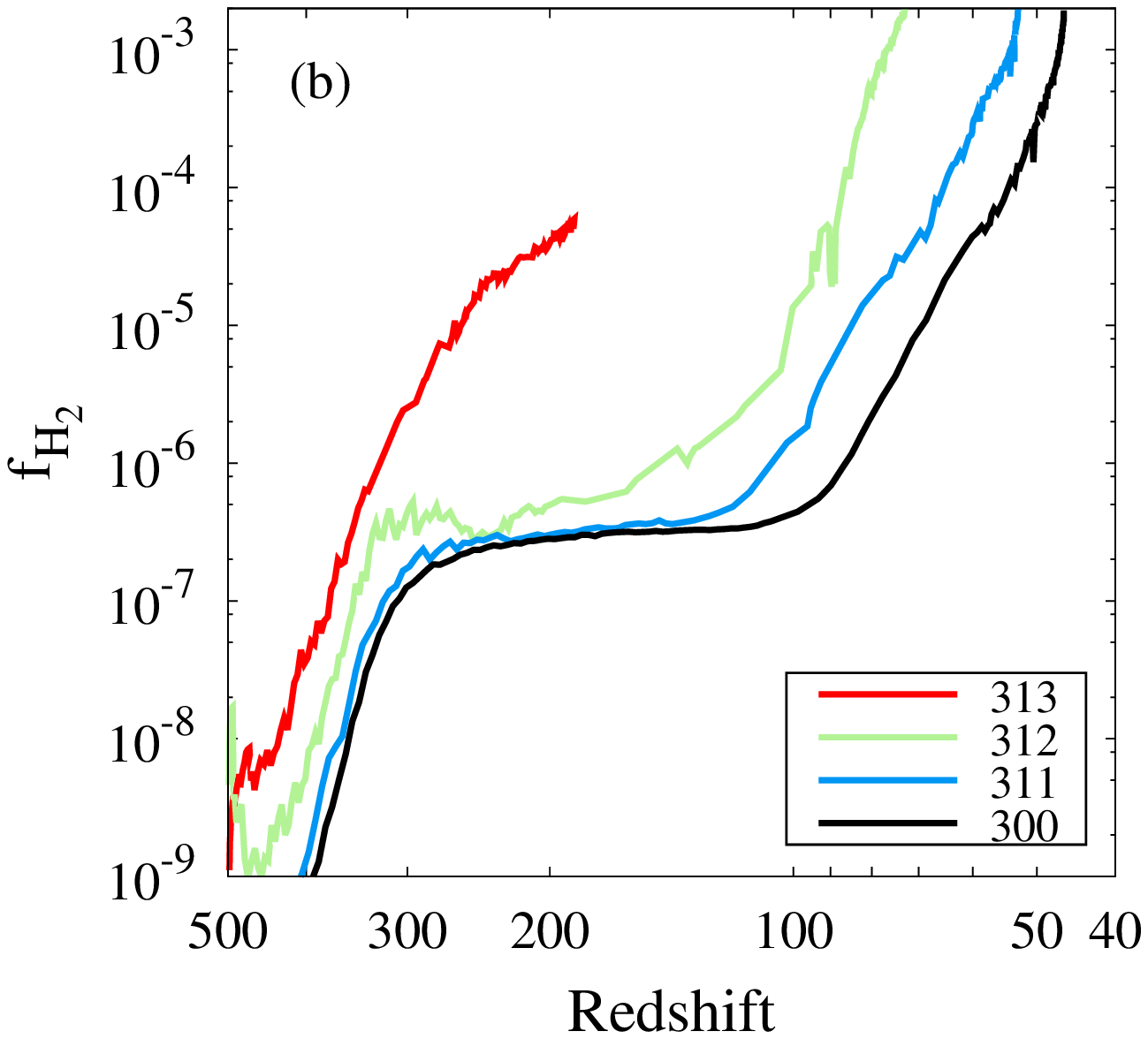}
\end{tabular}}
\caption{
Time evolution of gas density (Panel a) and H$_2$ molecule fraction (Panel b) of the densest gas elements for cases with $\sigma_8 = 2.5$ until $\nhcen = 10^6~\cc$. 
The lines with different colors represent the four PPS models in Table~\ref{tab:list}.
}
\label{fig:PPS-Ncent_z-XXX}
\end{center}
\end{figure*}

\section{Results}
\label{sec:res}

\subsection{Major Effects}

We find major effects of the small-scale power spectrum on the properties of host halos and on the star formation within them.
 Figure~\ref{fig:2dmap} shows the local conditions around the formation site of the primordial star. 
The top panels are for the case with normal scale-invariant PPS (ID 300). 
The primordial star-forming cloud is formed at $z_{\rm form} = 46$ inside a dark matter mini-halo that has a virial radius of $R_{\rm vir} \simeq 19$~pc. 
The bottom panels are for the case with an extremely enhanced PPS (ID 313). 
The gas cloud is formed earlier at $z_{\rm form} = 186$ inside a more compact mini-halo with $R_{\rm vir} \simeq 3.6$~pc. 
Figure~\ref{fig:2dmap}(a) shows the density distribution of dark matter component within a proper 100~pc on a side. 
Note that the mean density is higher in the bottom case than in the top because of the cosmic expansion, as $\rho \propto (1+z)^3$. 
The number of small mass dark matter clumps is significantly larger in the bottom panel, reflecting the enhanced initial density perturbations at the corresponding length scales (e.g., Figs.~\ref{fig:PPS} and \ref{fig:HMF}). 
Interestingly, we find a secondary star-forming cloud formed in the vicinity of the first one within several 10~pc (Fig.~\ref{fig:2dmap}b).
Such a close pair of primordial stars likely affect each other through radiative, dynamical, and chemical feedback effects. 
In the top panels, we also find a neighboring mini-halo 
but it has not grown sufficiently 
to host a dense star-forming cloud (see also the temperature 
distribution in Panel c).

The host mini-halos forming at very high redshifts are generally more compact and have larger dark matter densities, whereas the characteristic density of the gas cloud is determined largely by molecular hydrogen cooling.
Consequently, the baryonic component becomes dynamically dominant only after the gas density exceeds $\simeq 10^5~\cc$ in very early halos.

\subsection{Formation Epoch and H$_2$ Formation}

Figure~\ref{fig:PPS-Ncent_z-XXX}(a) shows the density evolution of the gas cloud's core as a function of redshift for the cases with $\sigma_8 = 2.5$. 
With more enhanced power (larger $m_{\rm s}$), the clouds collapse earlier.
In the extreme case with $m_{\rm s} = 2.5$ (red line), the cloud collapses at $z \sim 186$. 
Then H$_2$ formation mainly proceeds via the H$_2^+$ channel,
but the photo-dissociation by CMB prevents H$_2$ formation.
Hence the molecular fraction remains low for ID 313 until three-body H$_2$ formation operates at very high densities (Fig.~\ref{fig:PPS-Ncent_z-XXX}b). 
The formation of H$_2$ molecules critically affects the thermal evolution, prestellar collapse, and the subsequent accretion process onto the central protostar.

\begin{figure*}
\begin{center}
\resizebox{15cm}{!}{\begin{tabular}{cc}
\includegraphics[clip,scale=1]{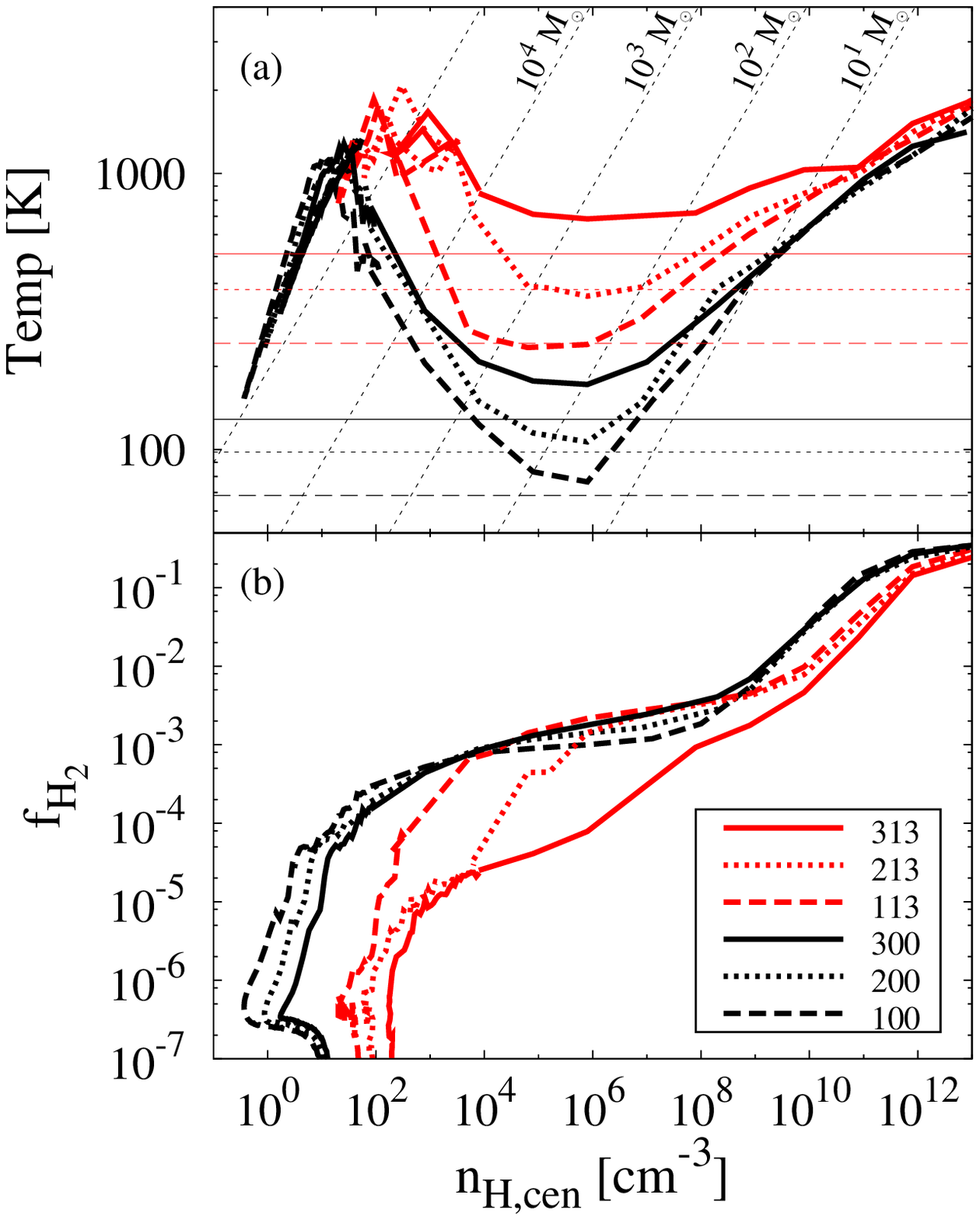}&
\includegraphics[clip,scale=1]{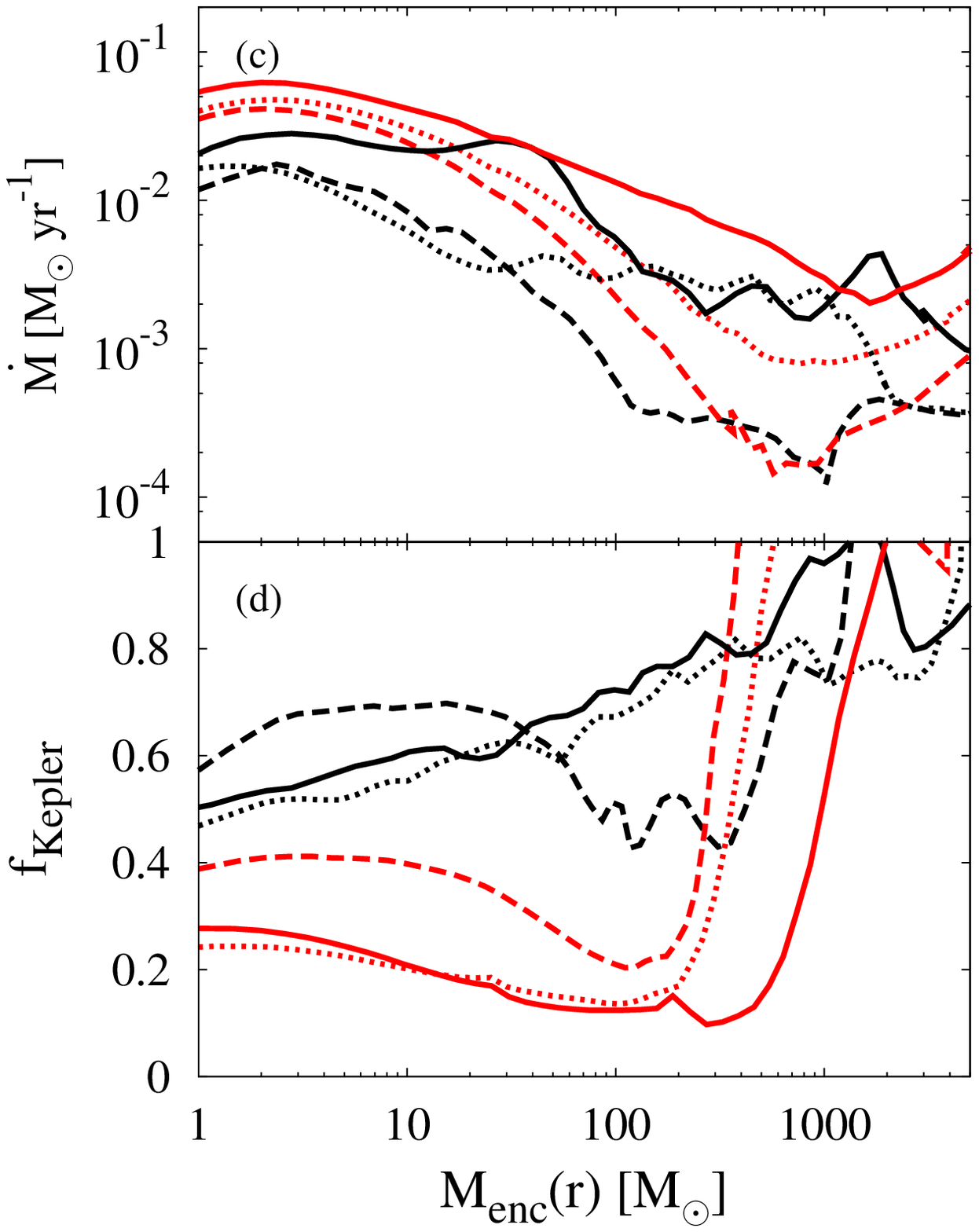}
\end{tabular}}
\caption{
The evolution of temperature (Panel a) and H$_2$ fraction (Panel b) 
at the center of the cloud.
The horizontal lines in Panel (a) indicate the CMB temperature, $T_{\rm CMB}(z) = 2.73~(1+z)$, at the respective formation epoch $z_{\rm form}$ listed in Table~\ref{tab:list}. 
The dotted lines show the Jeans mass for $10 - 10^5~\msun$ in the density-temperature plane.
Panels (c) and (d) display the profiles of the instantaneous gas infall rate and the degree of rotational support ($f_{\rm Kepler} = v_{\rm rot} / v_{\rm Kepler}$) as a function of the enclosed mass. 
The profiles are calculated when the cloud core density reaches $\nhcen = 10^{13}~\cc$.
}
\label{fig:Profiles}
\end{center}
\end{figure*}

\subsection{Thermal and Dynamical Properties of Clouds}

Figure~\ref{fig:Profiles} shows the evolution and the properties of collapsing clouds for the six cases for which direct RHD simulations are performed. 
We use black lines for models with the standard power-law spectrum ($m_{\rm s} = n_{\rm s}$) and we use red lines for models with enhanced small scale power ($m_{\rm s} > 1$). 
The thermal evolution in the standard model is well understood and can be described as follows (see Panel a). 
The temperature first increases due to gravitational contraction but then decreases owing to H$_2$ cooling, when the molecular fraction reaches the critical value that is needed to cool the gas within a Hubble time $f_{\rm H_2} \simeq 5 \times 10^{-4}$ \citep[see Fig.~\ref{fig:Profiles}b, e.g.,][]{tegmark97}. 
Note that the minimum temperature is floored by $T_{\rm CMB}$ at high redshifts.
The gas temperature starts increasing again when the compressional heating overcomes the radiative cooling, leading to the formation of a protostar.

In our models with enhanced small scale power (red lines), the cloud evolves on a higher temperature track because of the lower H$_2$ fraction (Panel b).
With larger $m_{\rm s}$, the density at which the H$_2$ fraction reaches the critical value for cooling is larger.
Interestingly, for ID 313 (red solid line), the H$_2$ fraction does not reach the critical value until three-body H$_2$ formation occurs at $\nhcen > 10^8~\cc$. 
In this case, the temperature at the onset of run-away collapse is 700~K, much higher than that found in the usual primordial star formation. 
One can naively expect that protostars formed in such a 'hot' cloud will grow very rapidly.

The stellar mass can be estimated from Eq.~(\ref{eq:MIII_dMdt-Jeans}).
Effectively, the mass accretion history determines the protostellar evolution and the strength of radiative feedback that halts the accretion of surrounding medium.
Figure~\ref{fig:Profiles}(c) shows the gas infall rate as a function of the enclosed gas mass. 
On average, the gas infall rate systematically increases at higher redshift (see the difference from long-dashed to solid lines). 
However, the gas infall rate does not monotonically increase with $m_{\rm s}$.
This is because the host mini-halos of the gas clouds are formed at earlier epochs but contain less amount of gas.

Figure~\ref{fig:Profiles}(d) shows the ratio of rotational speed to the Keplerian speed. 
Interestingly the gas clouds formed at earlier epochs spin significantly more slowly in the models with enhanced power.
This may reflect the effective redistribution of the (orbital) angular momenta brought in by numerous minor mergers. 
Stars formed in a gas cloud with less spin tend to be more massive because of relatively rapid spherical accretion onto the protostar \citep{hirano14}.

\subsection{Dependence on Model Parameters}

It is important to study the effect of the pivot scale $k_{\rm p}$ of the PPS on gas cloud formation.
There is an important physical scale that characterizes primordial star formation. 
Let us denote the characteristic wavenumber $k_{\rm halo}$ for a mini-halo that has a physical size $R_{\rm halo}$ so that $k_{\rm halo} \sim 2\pi / R_{\rm halo}$.
When $k_{\rm p}$ is smaller than $k_{\rm halo}$, the enhanced power generates more structure of mini-halo sizes.

In the fiducial case of ID = 100, the host halo has a virial mass of $\sim 2 \times 10^5~ \msun$. 
The corresponding wavenumber is $k_{\rm halo} = 2 \pi / R_{\rm halo} \sim 500~h~{\rm Mpc}^{-1}$. 
When $k_{\rm p}$ is comparable to $k_{\rm halo}$, as in the cases with $k_{\rm p} = 300$ and $500~h~{\rm Mpc}^{-1}$, increasing $m_{\rm s}$ does not significantly affect $z_{\rm form}$ and $M_{\rm vir}$ (see Table \ref{tab:list}).
This is in direct contrast to the trend found when $k_{\rm p} < k_{\rm halo}$; increasing $m_{\rm s}$ actually resulted in {\it larger} $z_{\rm form}$ and {\it smaller} $M_{\rm vir}$.
This can be understood as follows. 
The formation epoch is largely determined by the amplitude of large-scale density perturbations whereas the enhanced power promotes formation of small mass halos.
Therefore, when $k_{\rm p}$ is small, density perturbations at all the relevant length scales are enhanced. 
The mini-halos are then formed early, and are physically compact.

It is worth considering possible halo-to-halo variation. Our understanding so far is based on a limited number of samples. 
\cite{hirano14,hirano15} show substantial differences in the properties of star-forming clouds even with similar masses and similar formation epochs that are in turn determined largely by the cosmological parameters.
In principle, using a much large number of halos  could show the general trend and variations more clearly, but we have focused on qualitative differences caused by the shape of PPS.

\section{Discussion}
\label{sec:dis}

We have studied how the shape of the primordial power spectrum at small scales affects early structure formation. 
For a significantly enhanced PPS, star-forming gas clouds are formed at extremely early epochs of $z > 100$ (ID 213 and 313), when the usual gas-phase reaction of molecular hydrogen formation does not operate.
At $100 < z < 300$, hydrogen molecules are formed via the H$_2^+$ channel and thus the gas can still cool and condense to form primordial stars.
The formed stars have masses as large as $\sim 300~\msun$.

At even earlier epochs of $z > 300$, there is no efficient formation channel of H$_2$ and the possible path to star formation is through cooling by atomic hydrogen.
If the highly enhanced density perturbations, as explored in the present paper, grow quickly to form primordial gas clouds in the very early universe, an extremely massive star might be formed in a similar manner to the so-called direct collapse black hole formation model \citep[e.g.,][]{inayoshi14}.
Such very massive stars gravitationally collapse to form remnant massive black hole at the end of their lives, leaving promising ``seeds'' for the formation of the recently discovered supermassive black hole \citep[SMBH;][]{wu15}.

There is an additional generic physical process that can significantly affect the structure formation in the early universe; the relative motions between dark matter and baryons caused by acoustic oscillations at the cosmic re-combination era \citep{tseliakhovich10}. 
The relative streaming motions prevent the gas contraction into dark matter mini-halo and delay the star formation, and thus possibly change the physical condition of the first star formation \citep[see][for recent review]{fialkov14}.
The effect is important at the very high redshifts considered in this paper, since the root-mean-square streaming velocity scales with redshift as $\propto (1+z)$. 
We have run several additional simulations and confirmed that, under significant streaming motions, primordial gas clouds are formed later in large host halos
than presented in the present paper.
There is an interesting possibility that streaming motions delay the first star formation to the extent that gas collapse is suppressed in mini-halos and more massive halos eventually host very massive stars, which can become the seed of SMBH \citep{tanaka14}.
The blue-tilted primordial power-spectral can effectively increase the number of massive halos in the early universe.
Further results from simulations with the streaming motions will be presented elsewhere (Hirano et al., in preparation).

Observationally, an independent probe of the PPS at small length scales might be provided by the existence of small mass dark halos in and around galaxies. 
Following the analytical model of \cite{zhao05}, we estimate that an extremely large number of halos with masses of $\sim 1~\msun$ are formed for $m_{\rm s} = 1.5$ and $k_{\rm p} = 100$~Mpc$^{-1}$.
The typical formation epoch of such solar-mass halos is $z \sim 300$, and thus the halos may remain as very compact objects. 
However, even the dense and compact clumps can be destroyed by tidal stripping/disruption and encounter with stars in the Galaxy.
It is thus difficult to estimate the abundance of the mini- and micro-halos surviving until today, either by large scale simulation \citep[e.g.,][]{diemand05,green07} or by analytical calculation \citep{zhao05}. 
Nevertheless, if signatures of small mass dark halos are detected through, for instance, dark matter annihilation or direct detection experiments, one can obtain invaluable information on their origin, i.e., the small-scale power spectrum.

Currently, there are little direct probes of the small-scale primordial density fluctuations, and only model dependent constraints are available \citep{chluba12,berezinsky14,natarajan15}. 
We look forward to future CMB experiments such as PRISM\footnote{http://www.prism-mission.org/} and PIXIE \citep{kogut11} that can infer the formation epoch and the spectral energy distribution (hence roughly the typical mass) of the first stars through measurement of CMB spectral distortions and the detection of low level of early reionization at $l > 10$ \citep[e.g.,][]{ricotti05}.
Finally, James Webb Space Telescope \citep{windhorst09} and WFIRST observations \citep{spergel15} will probe the existence of quasars at even higher redshifts.
Understanding the nature of the first stars may thus provide a route to probe observationally the small-scale primordial power spectrum.

\acknowledgments 
We thank Jens Chluba and Carla Maria Coppola for helpful advices about 
chemistry implementations. 
We also thank Takashi Hosokawa, 
Kazuyuki Omukai and Teruaki Suyama for discussions and 
comments on the earliest star formation. 
NZ is grateful for the hospitality of Department of Astrophysical 
Sciences at Princeton University. NZ's visit was supported by 
the University of Tokyo-Princeton strategic partnership grant.
The numerical calculations were carried out on Cray XC30 and 
the general-purpose PC farm at Center for Computational Astrophysics, 
CfCA, of National Astronomical Observatory of Japan.
This work was supported by Grant-in-Aid for JSPS Fellows (SH).
and by the JSPS Grant-in-Aid for Scientific Research 25287050 (NY).
Portions of this research were conducted at the Jet Propulsion Laboratory, California Institute of Technology, operating under a contract with the National Aeronautics and Space Administration (NASA).

\bibliography{biblio}
\bibliographystyle{apj}

\appendix

\section{Reaction rate calculation of H$_2^+$ photo-dissociation}
\label{app:H2+}

The reaction rate calculation of H$_2^+$ photo-dissociation is highly uncertain. 
In our previous calculations \citep{hirano14,hirano15}, we adopt the fitting function of \cite{galli98} assuming the LTE-level population. 
The accurate reaction rate can be calculated by considering the H$_2^+$-level population \cite[][]{hirata06,coppola11b}. 
The time evolution of H$_2$ abundance at $z > 60$, which is the relevant epoch in our study, differs significantly depending on the adopted reaction rate \citep[see fig. 3a in][]{galli13}. 
We take a straightforward approach by comparing simulations with different reaction rates.
Fortunately, we find that the uncertainty in the reaction rate gives essentially no influence on the primordial star formation. 
Stars are formed in dense gas clouds in which a critical amount of hydrogen molecules are formed to overcome the compressional heating by their radiative cooling. 
When a gas cloud becomes gravitationally unstable after the so-called loitering phase, the H$_2$ fraction within the cloud reaches a threshold value that is independent on the treatment of H$_2^+$ photo-dissociation.
We thus adopt the LTE rate to produce the main results in the present paper.
In most cases, we find a constant ``floor'' of the H$_2$ fraction at $2 \times 10^{-7}$ owing to the H$_2$ formation via the H$_2^+$ channel (Fig.~\ref{fig:PPS-Ncent_z-XXX}b).

\end{document}